\documentclass{ws-procs975x65}

\begin{document}

\title{MAGNETIZATION OF COLOR-FLAVOR LOCKED MATTER}

\author{J. NORONHA$^*$}

\address{Frankfurt Institute for Advanced Studies, J.W. Goethe--Universit\"at,\\
Frankfurt am Main, D-60438, Germany\\
$^*$E-mail: noronha@fias.uni-frankfurt.de}

\author{I. A. SHOVKOVY}

\address{Department of Physics, Western Illinois University,\\
Macomb, IL 61455, USA\\
E-mail: I-Shovkovy@wiu.edu}

\begin{abstract}
We show that the magnetization in color-flavor locked superconductors can be
so strong that homogeneous quark matter becomes metastable for a wide range of
magnetic field values. This indicates that magnetic domains or other type of magnetic
inhomogeneities can be present in the quark cores of magnetars.
\end{abstract}


\bodymatter

\section{Introduction and Conclusions}\label{aba:sec1}

The cold and superdense inner cores of compact stars are likely to consist
of color superconducting quark matter. Assuming that the very strong surface
magnetic fields found in magnetars ($B \sim 10^{14}- 10^{16}$~G) can be
transmitted to the their inner cores, it is worth studying whether the effects
of strong magnetic fields on color superconductors are important for
understanding the physics of magnetars. In Ref.~\refcite{mcfljorge} the
effects of moderately strong magnetic fields on the properties of
color-flavor locked (CFL) superconductors (Ref.~\refcite{CFL}) were
studied numerically in the framework of a Nambu-Jona-Lasinio model.
It was shown that the ground state of 3-flavor quark matter undergoes
a continuous crossover from the CFL phase into the magnetic CFL
(mCFL) phase, which was initially introduced in Ref.~\refcite{cristina}.

\begin{figure}[!ht]
\centering
\epsfig{file=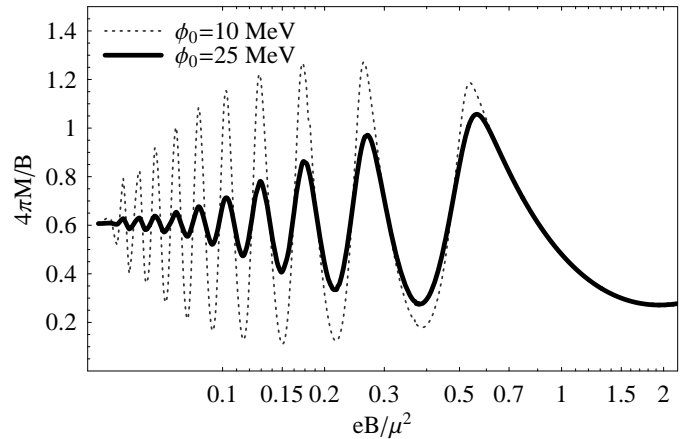,width=0.7\linewidth}
\caption{Ratio $4\pi M/B$ versus $eB/ \mu^2$ for two sets of parameters
that yield $\phi_{0}= 10$~MeV and $\phi_{0}= 25$~MeV.}
\label{figmag}
\end{figure}

The free parameters of the model (the diquark coupling constant and
the ultraviolet cutoff) were set to yield a CFL gap of either
$\phi_{0}= 10$~MeV or $\phi_{0}= 25$~MeV when the quark chemical
$\mu=500$ MeV and $B=0$. It was shown in Ref.~\refcite{mcfljorge}
that the mCFL gaps display de Haas-van Alphen oscillations with respect
to $eB/\mu^2$ (see also Ref.~\refcite{FukWar}). The magnetization of the
system $M$ is given by $M=(\partial \Gamma/\partial B)$, where $\Gamma$
is the one-loop quark contribution to the free energy evaluated at
the stationary point. Our numerical results for the magnetization are
presented in Fig.~\ref{figmag}.

It was shown in Ref.~\refcite{lattimer} that the magnetization of hadronic
matter is negligible even for magnetar conditions, i.e., $4\pi M/B\ll 1$ for
$B\lesssim 10^{19} G$. In the case of quark matter, however, the situation
is very different. In Fig.~\ref{figmag} we plotted $4\pi M/B$ versus
$eB/\mu^2$ for a mCFL superconductor. The magnetization is
significantly larger in this case and it displays de Haas-van Alphen oscillations
with very large amplitude. The large magnitude of these magnetic oscillations
create regions in which $(\partial H/\partial B)_{\mu}<0$, where $H=B-4\pi M$
is the magnetic field present in the outer layers of the star. These regions
correspond to unstable or metastable states where, depending on the
geometry of the system, a transition into a magnetic domain configuration
may occur. This could lead to a wealth of different physical phenomena.

For instance, successive phase transitions coming from discontinuous
changes of $B$ during the star's evolution can release a vast amount of
energy that would heat up the star, which would then cool down by for
example the emission of neutrinos. Therefore, sudden bursts of neutrinos
coming from magnetars with color superconducting cores even after the
deleptonization period could be expected. On the other hand, if a mixed
phase with microscopic domains of nonequal magnetizations is formed,
the relative size of domains with different magnetizations would change
with $H$ in order to keep the average induced magnetic field $B$
continuous. In either case, since the magnitude of the fields involved is
enormous the system could potentially release an immense amount of
energy. If observed, this type of phenomena could help to distinguish
magnetars with quark cores from their purely hadronic counterparts.

\end{document}